# Have Mysterious Topological Valley Currents Been Observed in Graphene Superlattices?

Stephan Roche,[1,2] Stephen R. Power,[3,4] Branislav K. Nikolić,[5] José Hugo García[1], and Antti-Pekka Jauho[6]

[1]Catalan Institute of Nanoscience and Nanotechnology, CSIC and The Barcelona Institute of Science and Technology,Campus UAB, Bellaterra, 08193 Barcelona (Cerdanyola del Vallès), Spain
[2]Institució Catalana de Recerca i Estudis Avançats, 08070 Barcelona, Spain
[3]School of Physics, Trinity College Dublin, Dublin 2, Ireland
[4]School of Physical Sciences, Dublin City University, Dublin 9, Ireland
[5]Department of Physics and Astronomy, University of Delaware, Newark, DE 19716, USA
[6]Center for Nanostructured Graphene, DTU Physics, Technical University of Denmark, Kongens Lyngby, DK-2800, Denmark

**Abstract:**
We provide a critical discussion concerning the claim of topological valley currents, driven by a global Berry curvature and valley Hall effect proposed in recent literature. After pointing out a major inconsistency of the theoretical scenario proposed to interpret giant nonlocal resistance, we discuss various possible alternative explanations and open directions of research to solve the mystery of nonlocal transport in graphene superlattices.

In recent years, the concept of *topological valley current* has been proposed to explain puzzling measurements of a giant nonlocal resistance in graphene/hexagonal boron-nitride (hBN)-based devices [1-7]. The weak interaction between graphene (bi-)layer aligned with commensurate registry onto hBN breaks inversion symmetry of graphene and results in the formation of a Moiré pattern which alters the electronic properties of graphene in a subtle fashion, with the formation of satellite Dirac cones at energies corresponding to the Moiré supercell lengths and the opening of a spatially non-uniform gap pattern. In nonlocal transport measurements, the surprising observation of a robust giant peak of the nonlocal resistance at the charge neutrality point has been initially assigned to an overall bulk energy gap which induces Berry curvature hot spots [8] just above and just and below the gap. The Berry curvature generates an anomalous velocity [8] and a valley-dependent current splitting, analogous to the intrinsic spin Hall effect [9] but with the valley degree of freedom playing the role of electron spin.

The original theoretical interpretation [10] of the experiments published in Ref. [1] essentially introduces a new type of topological effect in condensed matter, rather different from those for which the Nobel Prize in Physics was awarded in 2016 [11]. In this new picture, topological valley currents, which would carry the nonlocal signal observed in Ref. [1] in the center of the gap, are produced by Fermi sea states just beneath the gap, and not by gapless edge states [12] at the Fermi energy. This is unlike all other known cases [11] where "*unexpected topologically protected edge states recur again and again in connection with topological state of matter*" (quote taken from Ref. [11]). If correct, such an explanation [10] further suggests [13] a serious shortcoming of the Landauer-Büttiker formula, a cornerstone of mesoscopic [14] and nanoscale physics [15] which has successfully been used to interpret virtually all nonlocal quantum transport experiments (with [16] or without edge states [17]) since the 1990s. That is, the Landauer-Büttiker formula applied to multiterminal device from Ref. [1], described by the presumed gapped Hamiltonian from Ref. [10], predicts *zero* nonlocal resistance [13]. This would require theorists to either replace, or re-derive with possible additional terms, the Landauer-Büttiker formula thereby updating numerous standard textbooks [14,15].

Accordingly, the theoretical explanation in [10] has raised concerns given the apparent contradiction between measuring a Fermi surface signal inside a "gapped" device, but arguing for the effect carried by the states deep inside the Fermi sea: *"Naïvely, the lack of edge transport would lead one to conclude that topological currents cease to exist. If true, this would imply that the key manifestations, such as valley Hall conductivity and orbital magnetization, vanish in the gapped state. Here we argue that the opposite is true: the absence of conducting edge modes does not present an obstacle since valley currents can be transmitted by bulk states beneath the gap."* (quote from Ref. [10]). This was first questioned theoretically in Ref. [13]—where the emergence of nonlocal transport signal, as well as often measured experimentally [1,3,5,7] metallic-like longitudinal resistivity instead of expected large one for highly insulating state due to gapped band structure and low temperatures—was found to be consistent with the formation of non-topological edge currents. They are also resilient to disorder owing to their peculiar energy-momentum dispersion relation. Subsequent experiments [3,5,7] actually suggest a non-topological origin of nonlocal signals while emphasizing the importance of edge currents in this type of system, but without further clarification of underlying physical mechanisms at play. In fact, part of the disconnection between the bulk and device physics may originate from the definition of the valley operator, which requires an artificial separation of the Brillouin zone. It has been recently proposed that by considering the valley Hall effect as a manifestation of the orbital Hall effect [18], one can express the physics in terms of real observables while still producing robust edge-states. Although the orbital Hall effect does not require a broken inversion symmetry [19], such a connection opens the door for the debate of the proper definition of the valley current operator.

On the other hand, a recent work [20] demonstrates that valley Hall effects can indeed lead to an enhancement of the nonlocal resistance. However, the alternative mechanism at work here stems not from Berry curvature effects inside a global band gap, but from the spatially non-uniform gap profile that emerges in graphene/hBN heterostructures due to the Moiré pattern. The emergence of inhomogeneous gapped regions in a graphene sheet can be directly connected to a valley-dependent Hall effect which does not rely on Fermi sea contributions or a global gap. Instead, an extrinsic valley Hall effect is generated by local mass dots imprinted in the Moiré superstructure (see Figure 1 for illustration). This study, which can also be extended to bilayer graphene [21], reconciles the assignment of a Fermi surface contribution to the giant nonlocal resistance, which can be simulated brute-force by using textook [14,15] Landauer-Büttiker formula and can be related to a nonzero valley Hall conductivity that remains finite (although not quantized) at low energy. One also notes the reported possibility to generate quantum valley Hall effect by combining real magnetic fields with lattice deformation fields, which produces effective gauge fields, preserving time-reversal symmetry. The combined effects of gauge superposition and annihilation at K- and K' valleys generates a valley-polarized Hall current in one valley, revealed by a $e^2/h$ Hall conductivity plateau, with a concomitant dissipative valley-polarized current flow in the opposite valley [22].

These theoretical works send a clear message. By proper mass-term engineering, using specific substrates, controlled deposition of atomic clusters, imprinted molecular patterns, strain fields or assembly of tiny holes, the production of a bulk-driven valley Hall effects or even realization of quantum valley Hall effect can be tailored and fine-tuned, enabling further possibilities for manipulation of the valley degree of freedom of Dirac matter in the larger context of quantum materials [23].

In 2021, some of the authors of Ref[1] published a new manuscript in Nature entitled "*Long-range nontopological edge currents in charge-neutral graphene*" [24], in which they argue the potential predominant role of "edge (current) accumulation" in analysing nonlocal transport signals in absence or presence of external magnetic field, and finally conclude that "*The observation here of long-range edge currents not protected by topology, but nevertheless robust and coexisting with the bulk conduction, calls for careful re-examination of some of the reported nonlocal transport phenomena.*"

A direct imaging of the formation of valley-polarized edge currents, while concurrently measuring the same nonlocal resistance as observed earlier in Ref. [1,5,7], could actually provide the sought-after *smoking gun* for conclusive determination of the nonlocal transport mechanism, as also discussed in other context of topological physics [12] or hydrodynamics flow in Dirac materials [25].

We also note that the debate overviewed above is confined to quantum transport phenomena of noninteracting quasiparticles. Additional quantum many-body effects, such as due to on-site Coulomb interaction [26] which could close the gap of van der Waals heterostructure including graphene/hBN [27], are beyond either the standard Kubo formula used in Refs. [1,8,10,13] or the Landauer-Büttiker formula used in Ref. [13]. Their inclusion into a quantum transport formalism capable to model multiterminal devices would require major new theoretical advances [28].

The final take home message is that while the concept of topological valley currents to explain nonlocal transport measurements seems flawed and incorrect, several solid alternative explanations of experimental data have been proposed [13,20,21], while other directions including the role of orbital Hall effect or the contribution of nonequilibrium phenomena (far beyond the limited reach of semiclassical transport phenomenology) are also worth exploring. Beyond the advance of our comprehension in topological physics in graphene superlattices and related structures, such research could also further unravel novel opportunities for realizing quantum information manipulation in two-dimensional materials, van der Waals heterostructures and Moiré quantum matter [23]. We note that a complementary discussion, including aspects on the possible shortcomings of the experimental measurements has been published in Ref. [29].

## **Acknowledgments**

S.R.P. acknowledges funding from the Irish Research Council under the Laureate awards program. J.H.G. and S.R. acknowledge funding from the European Union Seventh Framework Programme under Grant No. 881603 (Graphene Flagship). ICN2 is funded by the CERCA Programme/Generalitat de Catalunya and supported by the Severo Ochoa programme (MINECO Grant. No. SEV-2017-0706). The Center for Nanostructured Graphene is sponsored by the Danish National Research Foundation, Project No. DNRF103. B. K. N. was supported by the US National Science Foundation (NSF) through the University of Delaware Materials Research Science and Engineering Center DMR-2011824.

## **BIBLIOGRAPHY**

[1] R. Gorbachev J. C. W. Song, G. L. Yu, A. V. Kretinin, F. Withers, Y. Cao, A. Mishchenko, I. V. Grigorieva, K. S. Novoselov, L. S. Levitov, and A. K. Geim, *Detecting topological valley currents in graphene superlattices,* Science **346**, 448 (2014).

[2] M. Sui, G. Chen, L. Ma, W.-Y. Shan, D. Tian, K. Watanabe, T. Taniguchi, X. Jin, W. Yao, D. Xiao, and Y. Zhang *Gate-tunable topological valley transport in bilayer graphene*, Nat. Phys. **11**, 1027 (2015).

[3] M. J. Zhu, A. V. Kretinin, M. D. Thompson, D. A. Bandurin, S. Hu, G. L. Yu, J. Birkbeck, A. Mishchenko, I. J. Vera-Marun, K. Watanabe, T. Taniguchi, M. Polini, J. R. Prance, K. S. Novoselov, A. K. Geim, and M. Ben Shalom, *Edge currents shunt the insulating bulk in gapped graphene*, Nat. Commun. **8**, 14552 (2017).

[4] Y. Shimazaki, M. Yamamoto, I. V. Borzenets, K. Watanabe, T. Taniguchi, and S. Tarucha *Generation and detection of pure valley current by electrically induced Berry curvature in bilayer graphene*, Nat. Phys. **11**, 1032 (2015).

[5] K. Komatsu, Y. Morita, E. Watanabe, D. Tsuya, K. Watanabe, T. Taniguchi, and S. Moriyama *Observation of the quantum valley Hall state in ballistic graphene superlattices*, Sci. Adv. **4**, eaaq0194 (2018)

[6] K. Endo, K. Komatsu, T. Iwasaki, E. Watanabe, D. Tsuya, K. Watanabe, T. Taniguchi, Y. Noguchi, Y. Wakayama, Y. Morita, and S. Moriyama, *Topological valley currents in bilayer graphene/hexagonal boron nitride superlattices,* Appl. Phys. Lett. **114**, 243105 (2019).

[7] Y. Li, M. Amado, T. Hyart, G. P. Mazur, and J. W. A. Robinson, *Topological valley currents via ballistic edge modes in graphene superlattices near the primary Dirac point*, Commun. Phys. **3**, 224 (2020).

[8] D. Xiao, W. Yao, and Q. Niu, *Valley-contrasting physics in graphene: Magnetic moment and topological transport*, Phys. Rev. Lett. **99**, 236809 (2007).

[9] J. Sinova, S. O. Valenzuela, J. Wunderlich, C. H. Back, and T. Jungwirth, *Spin Hall effects*, Rev. Mod. Phys. **87**, 1213 (2015).

[10] Y. D. Lensky, J. C. W. Song, P. Samutpraphoot, and L. S. Levitov, *Topological valley currents in gapped Dirac materials*, Phys. Rev. Lett. **114**, 256601 (2015).

[11] F. Duncan M. Haldane, *Nobel Lecture: Topological quantum matter*, Rev. Mod. Phys. **89**, 040502 (2017).

[12] U. Bajpai, M. J. H. Ku, and B. K. Nikolić, *Robustness of quantized transport through edge states of finite length: Imaging current density in Floquet topological versus quantum spin and anomalous Hall insulators*, Phys. Rev. Res. **2**, 033438 (2020).

[13] J. M. Marmolejo-Tejada, J. H. García, M. Petrović, P.-H. Chang, X.-L. Sheng, A. Cresti, P. Plecháč, S. Roche, and B. K. Nikolić, *Deciphering the origin of nonlocal resistance in multiterminal graphene on hexagonal-boron-nitride with ab initio quantum transport: Fermi surface edge currents rather than Fermi sea topological valley currents*, J. Phys. Mater. **1**, 0150061 (2018).

[14] S. Datta, *Electronic Transport in Mesoscopic Systems* (Cambridge University Press, Cambridge, 1995).

[15] Y. V. Nazarov and Y. M. Blanter, *Quantum Transport: Introduction to Nanoscience* (Cambridge University Press, Cambridge, 2009).

[16] P. L. McEuen, A. Szafer, C. A. Richter, B. W. Alphenaar, J. K. Jain, A. D. Stone, R. G. Wheeler, and R. N. Sacks, *New resistivity for high-mobility quantum Hall conductors*, Phys. Rev. Lett. **64**, 2062 (1990).

[17] D. V. Tuan, J. M. Marmolejo-Tejada, X. Waintal, B. K. Nikolić, and S. Roche, *Spin Hall effect and origins of nonlocal resistance in adatom-decorated graphene*, Phys. Rev. Lett. **117**, 176602 (2016).

[18] S. Bhowal and G. Vignale, *Orbital Hall effect as an alternative to valley Hall effect in gapped graphene*, Phys. Rev. B 103, 195309 (2021)

[19] T. P. Cysne, M. Costa, L. M. Canonico, M. Buongiorno Nardelli, R. B. Muniz, and T. G. Rappoport, *Disentangling Orbital and Valley Hall Effects in Bilayers of Transition Metal Dichalcogenides*, Phys. Rev. Lett. 126, 056601 (2021)

[20] T. Aktor, J.H. Garcia, S. Roche, A.-P. Jauho and S. R. Power, *Valley Hall effect and nonlocal resistance in locally gapped graphene*, Phys. Rev. B **103**, 115406 (2021).

[21] F. Solomon and S. R. Power, *Valley current generation using biased bilayer graphene dots*, Phys. Rev. B 103, 235435 (2021)

[22] M Settnes, J.H. García, S Roche, *Valley-Polarized Quantum Transport Generated by Gauge Fields in Graphene´* 2D Materials 4 (3) (2017)

[23] F. Giustino, J. H. Lee, F. Trier, M. Bibes, S. M. Winter, R. Valentí, Y.-W. Son, L. Taillefer, C. Heil, A. I. Figueroa, B. Plaçais, Q. S. Wu, O. V. Yazyev, E. P. A. M. Bakkers, J. Nygård, P. Forn-Diaz, S. De Franceschi, J. W. McIver, L. E. F. Foa Torres, T. Low, A. Kumar, R. Galceran, S. O. Valenzuela, M. V. Costache, A. Manchon, E.-A. Kim, G. R. Schleder, A. Fazzio, and S. Roche, *The 2021 quantum materials roadmap*, J. Phys. Mater. **3**, 042006 (2021).

[24] A. Aharon-Steinberg, A. Marguerite, D. J. Perello, K. Bagani, T. Holder, Y. Myasoedov, L. S. Levitov, A. K. Geim & E. Zeldov, "*Long-range nontopological edge currents in charge-neutral graphene*", Nature 593, 528–534 (2021)

[25] M. J. H. Ku, T. X. Zhou, Q. Li, Y. J. Shin, J. K. Shi, C. Burch, L. E. Anderson, A. T. Pierce, Y. Xie, A. Hamo, U. Vool, H. Zhang, F. Casola, T. Taniguchi, K. Watanabe, M. M. Fogler, P. Kim, A. Yacoby, and R. L. Walsworth, *Imaging viscous flow of the Dirac fluid in graphene*, Nature **583**, 537 (2020).

[26] H.-K. Tang, J. N. Leaw, J. N. B. Rodrigues, I. F. Herbut, P. Sengupta, F. F. Assaad, and S. Adam, *The role of electron-electron interactions in two-dimensional Dirac fermions,* Science **361**, 570 (2018).

[27] J.-R. Xu, Z.-Y. Song, C.-G. Yuan, and Y.-Z. Zhang, *Interaction-induced metallic state in graphene on hexagonal boron nitride*, Phys. Rev. B **94**, 195103 (2016).

[28] R. E. V. Profumo, C. Groth, L. Messio, O. Parcollet, and X. Waintal, *Quantum Monte Carlo for correlated out-of-equilibrium nanoelectronic devices*, Phys. Rev. B **91**, 245154 (2015).

[29] L.E.F. Foa-Torres and S.O. Valenzuela, "*A valley of opportunities*", Physics World 43-46 (2021)

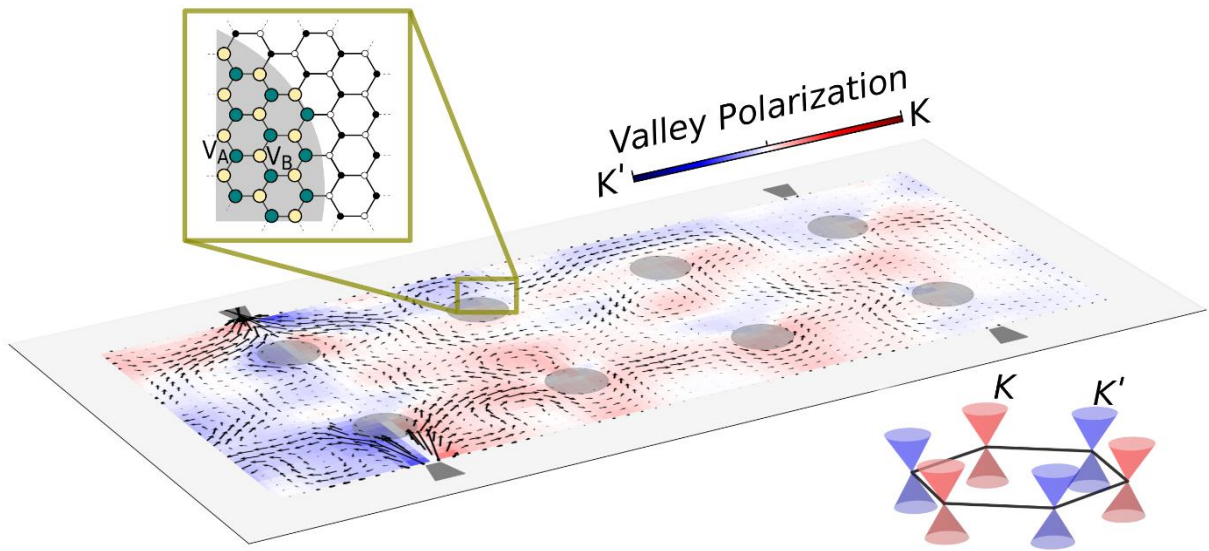

Figure 1: Overall picture of the formation of valley polarized currents in the mass-dot structure as analysed in Ref. [18]